\documentclass[preprint,showpacs,preprintnumbers,amsmath,amssymb, superscriptaddress,longbibliography,nofootinbib]{revtex4-1}
\usepackage{graphicx} 
\usepackage{xcolor}
\usepackage{amsmath}
 \usepackage{hyperref}
\usepackage[top = 2cm, bottom = 2cm, right = 2cm, left =2cm]{geometry}

\begin{document}

\title{A new representation of vacuum Lovelock solutions in $d = 2N+1$ dimensions: Black holes with an integrable singularity and regular black holes.}

\author{Milko Estrada }
\email{milko.estrada@gmail.com}
\affiliation{Facultad de Ingeniería y Empresa, Universidad Católica Silva Henríquez, Chile}

\date{\today}

\begin{abstract}
In recent years, black hole (BH) solutions with an integrable singularity have garnered significant attention as alternatives to regular black holes (RBH). In these models, similarly to RBHs, an object would not undergo spaghettification when approaching the radial origin. Instead of the potentially unstable de Sitter core present in RBHs, an integrable singularity emerges where the Ricci scalar diverges while its volume integral remains finite. However, the construction of both RBH solutions and BHs with an integrable singularity typically requires the inclusion of specific forms of matter in the energy-momentum tensor. We demonstrate that, from a geometric perspective in the absence of matter, vacuum solutions in Lovelock gravity in $d=2N+1$ dimensions can be represented as vacuum BHs with an integrable singularity in Einstein-Gauss-Bonnet theory for $d=5$ and in cubic gravity for $d=7$. Meanwhile, the vacuum solution in quartic gravity in $d=9$ is described as a vacuum RBH with a nontrivial hyperboloidal cross-section. For all the aforementioned cases, we have determined the conditions that the parameters in the solutions must satisfy. Remarkably, in all discussed cases, there is no presence of an internal horizon near a potentially unstable de Sitter core.
\end{abstract}

\maketitle

\section{Introduction}

The recent detection of gravitational waves from the collision of two rotating black holes \cite{LIGOScientific:2016aoc,LIGOScientific:2017ycc} has established them as some of the most exciting and intriguing objects in gravitation. Black hole (BH) solutions often exhibit a central singularity where metric and curvature invariants diverge. It is well known that the infinite tidal forces near a black hole's singularity can lead to extreme stretching of an object, a phenomenon commonly referred to as spaghettification. To address this issue, the standard approach involves defining a specific type of matter within the energy-momentum tensor, resulting in the formation of regular black holes (RBHs). RBHs are characterized by having finite values for curvature invariants and tidal forces throughout their spacetime.

It is well known the fact that RBHs solution usually incorporate the presence of an inner horizon associated with a de Sitter core close to the origin. In this connection, in recent years, there has been growing interest in determining whether the presence of an inner horizon is inherently unstable. For example, reference \cite{Carballo-Rubio:2022pzu} asserts that instability in the cores of regular black holes is inevitable because mass inflation instability is crucial for regular black holes with astrophysical significance. See also \cite{Ovalle:2023vvu} where it is claimed that predictability ceases after crossing the inner horizon.  However, this remains an unresolved issue from a theoretical standpoint. Conversely, reference \cite{Bonanno:2022jjp} argues that semiclassical effects due to Hawking radiation might mitigate the instability associated with the inner horizon, potentially stabilizing the existence of a de Sitter core. See also \cite{Ghosh:2022gka}.

As an alternative to RBH solutions, reference \cite{Lukash:2013ts} presents a gravitational scenario in which a particle would not be destroyed when approaching the origin of the radial coordinate. This methodology introduces specific types of matter sources such that, near the origin, an {\it integrable singularity} exists instead of a potentially unstable de Sitter core associated with the presence of an inner horizon. An integrable singularity is characterized by the divergence of the 
 Ricci scalar, while their volume integrals remain finite. Thus, the reference suggests that the presence of a central singularity results in finite tidal forces along the world line of the matter flow, preventing the destruction of objects as they approach the central integrable singularity. See the recent study of reference \cite{Casadio:2023iqt}, which offers a quantum interpretation of the existence of an integrable singularity. In this context, reference \cite{Estrada:2023dcj} recently outlined the characteristics that $d$-dimensional matter sources must have in order to construct a BH with no inner horizon, featuring a central integrable singularity instead of an potential unstable de Sitter core. It was demonstrated that such a central singularity is associated with finite tidal forces, the extendability of radial geodesics, and the weak nature of the singularity, as established by Tipler's extension theorem from reference \cite{Nolan:2000rn}. In this way, the elimination of the inner horizon prevents potential instabilities, and moreover, an object would not undergo spaghettification when approaching the central singularity. 

On the other hand, although the detection of gravitational waves \cite{LIGOScientific:2016aoc,LIGOScientific:2017ycc} has further established General Relativity (GR) as a correct theory of gravity at large scales, at quantum scales, it is well-known that General Relativity (GR) is not a complete theory, partly because it is non-renormalizable. This leads to the generation of higher curvature invariants in the action, which are required for renormalization \cite{Kuntz:2019qcf}. In this regard, as indicated in reference \cite{Hennigar:2017ego}, higher-order interactions are generally expected to appear in the low-energy effective action of the theory that provides ultraviolet (UV) completion to General Relativity.

One of the theories that considers the presence of high-curvature terms is Lovelock gravity. The Lagrangian of Lovelock gravity incorporates higher-curvature terms as corrections to the Einstein-Hilbert action \cite{Lovelock:1971yv}. Each correction can be viewed as a power of order $N$ with respect to the Riemann tensor. It is important to note that for dimensions $d \geq 2N + 1$, each correction contributes to the equations of motion. For even dimensions $d = 2N$, it corresponds to a topological invariant. Thus, we can observe that this theory admits the presence of extra dimensions and coincides with General Relativity (GR) when $d=4$. Furthermore, Lovelock's theories adhere to the fundamental principles of General Relativity; for instance, their equations of motion are of second order. Notably, the specific case of Lovelock gravity, known as Einstein-Gauss-Bonnet (EGB) theory, has gained attention in recent years for its applications in inflationary theories and has been compared to the results from GW170817 \cite{Odintsov:2020zkl}. See also \cite{Chakraborty:2016qbw,Chowdhury:2024uzd}

As mentioned earlier, the usual approach to constructing RBH solutions is based on the introduction of specific forms of matter in the energy-momentum tensor. In contrast to the latter, reference \cite{Bueno:2024dgm} recently demonstrated that, for a higher curvature theory corresponding to Quasi Topological Gravity (QTG) \cite{Myers:2010ru,Hennigar:2017ego} in $d \geq 5$, it is possible to construct vacuum RBH solutions purely from a geometric perspective. Specifically, this reference noted that in the vicinity of the radial origin $r = 0$, the highest order contribution of order $N$ dominates, and the solution near $r = 0$ is approximately given by:
\begin{equation} \label{SolucionBueno}
    \mu \approx 1 -\left( C_1 \right)^{1/N} r^{2 - (d-1)/N} + \cdots
\end{equation}
where $C_1$ is a constant. Thus, as $N \to \infty$, i.e., with the inclusion of an infinite tower of higher curvature terms, the solution leads to regular vacuum solutions with the presence of an internal horizon close to de Sitter core. For this, an appropriate definition of the coupling constants is imposed, such that the equations of motion reduce to a geometric series. It is important to emphasize that QTG, unlike Lovelock gravities, admits non-vanishing solutions for $d < 2N + 1$ when $d \geq 5$. Consequently, the fact that $N \to \infty$ makes it impossible to incorporate this infinite tower in Lovelock theories where $d \geq 2N + 1$. Therefore, this methodology is limited to applications in lower dimensions within QTG. For additional recent applications of this methodology, see references \cite{Konoplya:2024hfg,Konoplya:2024kih}. It is worth mentioning that for higher dimensions, $d \geq 2n+1$, QTG and Lovelock gravity exhibit similar structural features. 

Observing Equation \ref{SolucionBueno}, it seems that for $d = 2N + 1$, the core is non-divergent near the origin, but the Ricci scalar would diverge due to the absence of a Minkowski or (A)dS core, suggesting that we could have an integrable singularity. In this work, we will provide a new interpretation for vacuum black hole solutions in Lovelock gravity in $d = 2N + 1$ dimensions. The vacuum Einstein Gauss Bonnet (EGB) in $d = 5$, and the vacuum cubic solution in $d = 7$ will be represented as a black hole with an integrable singularity, while the vacuum quartic solution in $d = 9$ will be represented as a regular black hole with a nontrivial hyperboloidal cross-section. This contrasts with what is typically found in the literature, as both RBH solutions and black holes with an integrable singularity usually require the introduction of specific forms of matter in the energy-momentum tensor. In all the mentioned cases, there is an absence of a potentially unstable internal horizon. In Section \ref{Criterios}, we will establish the criteria for a vacuum Lovelock solution in $d = 2N + 1$ dimensions to represent an integrable singularity, without the presence of a potentially unstable inner horizon, where the geodesics can be extended to the singularity, and the singularity is of weak nature. Subsequently, we will write the Lovelock action as a finite geometric sum, thereby truncating the infinite tower from reference \cite{Bueno:2024dgm}. We will study the constraints that must be satisfied by the coupling constant $\alpha$ and the mass parameter $m$ in $d = 5$ in EGB theory and in $d = 7$ in cubic gravity for the solution to represent a vacuum black hole with an integrable singularity with the aforementioned characteristics. Finally, we will provide the solution in quartic gravity with a traversable hyperboloid section, which we will interpret as a vacuum regular black hole without the presence of an inner horizon.

\section{Equation of motion for vacuum Lovelock Black holes}

We begin by writing the generic Lovelock action:
\begin{equation} \label{AccionTopologica}
    I = \frac{1}{16\pi G} \int d^D x \sqrt{|g|} \left[ R + \sum_{n=2}^{n_{\text{max}}} \alpha_n Z_n \right]
\end{equation}
where $Z_n$ corresponds to the Quasitopological density of order $n$ with respect to the Riemann tensor \cite{Bueno:2024dgm}. It is worth mentioning that for $d \geq 2n+1$, $Z_n$ is similar to those corresponding to the Lovelock series \cite{Lovelock:1971yv}. The term  $R$ corresponds to the Ricci scalar. 

For $d \geq 2n+1$ (and consequently for our case of interest where $d = 2N+1$), the density $Z_n$ can be written in the following form $Z_n=\frac{1}{2^{n}}\, \delta_{\nu_{1}...\nu_{2n}}^{\mu_{1}...\mu_{2n}}\ R_{\mu_{1}\mu_{2}}^{\nu_{1}\nu_{2}}\cdots R_{\mu_{2n-1}\mu_{2n}}^{\nu_{2n-1}\nu_{2n}}$ where $\delta_{\mu_1 \dots \mu_n}^{\nu_1 \dots \nu_n}$ is the $n$-antisymmetric generalized Kronecker delta. From this formula, we can check that the $Z_2$ term is proportional to the Einstein-Gauss-Bonnet term $Z_2 \propto R^{\alpha\beta}_{\hspace{2ex}\mu\nu}R^{\mu\nu}_{\hspace{2ex}\alpha\beta}-4 R^{\alpha\nu}_{\hspace{2ex}\beta\nu} R^{\beta\mu}_{\hspace{2ex}\alpha\mu} + R^{\alpha\beta}_{\hspace{2ex}\alpha\beta}R^{\mu\nu}_{\hspace{2ex}\mu\nu}$. The $Z_3$ term is proportional to the third-order Lovelock term $
Z_3 \propto R^{3} - 12 R R_{\alpha \beta} R^{\alpha \beta} + 16 R_{\alpha \beta} R^{\alpha}{}_{\gamma} R^{\gamma \beta} + 24 R_{\alpha \beta} R_{\gamma \delta} R^{\alpha \beta \gamma \delta} + 3 R R_{\alpha \beta \gamma \delta} R^{\alpha \beta \gamma \delta} - 24 R_{\alpha \beta} R^{\alpha}{}_{\gamma \delta \epsilon} R^{\gamma \delta \epsilon} + 4 R_{\alpha \beta \gamma \delta} R^{\alpha \epsilon \gamma} R^{\delta \epsilon} - 8 R_{\alpha \beta \gamma}{}^{\delta} R^{\alpha \epsilon \gamma} R^{\beta}{}_{\delta \epsilon}$, and so on. The highest value that $n_{\text{max}}$ can take corresponds to the floor function of $d/2$, that is, $\lfloor d/2 \rfloor$.

The $d$-dimensional line element is:

\begin{equation} \label{metrica}
    ds^2=-\mu(r)dt^2+\mu(r)^{-1} dr^2+r^2 d\Sigma_\gamma. 
\end{equation}

The constant $\gamma$ can be normalized to $\pm 1$, $0$ by appropriately rescaling. Thus, the local geometry of $\Sigma_\gamma$ is a sphere, a plane, or a hyperboloid \cite{Aros:2000ij}:
\[
\Sigma_\gamma \text{ locally } =
\begin{cases}
S^{d-2}& \text{for } \gamma = 1 \\
T^{d-2} & \text{for } \gamma = 0 \\
H^{d-2} & \text{for } \gamma = -1.
\end{cases}
\]

In analogy to the fact that the variation of the action associated with the Ricci scalar gives rise to the Einstein equations in General Relativity, the variation of action \eqref{AccionTopologica} with respect to the metric, in the absence of the cosmological constant term of order zero, gives rise to the following equations of motion: \cite{Hennigar:2015mco,Camanho:2011rj,Bueno:2024dgm}: 

\begin{equation}
    \frac{d}{dr} \left( r^{d-1} h(\psi) \right) = 0 \Rightarrow h(\psi) = \frac{m}{r^{d-1}}
\end{equation}
where:
\begin{equation} \label{Suma}
    h(\psi) \equiv \psi + \sum_{n=2}^{n_{\text{max}}} \alpha_n \psi^n, 
\end{equation}
\begin{equation} \label{Psi}
 \psi \equiv \frac{\gamma - \mu(r)}{r^2}   
\end{equation}

\subsection{Our Assumptions for solutions with a finite series of higher-curvature terms}

For our case of interest, i.e., for $n_{\text{max}} = N = (d-1)/2$, with $d$ being odd and $d \geq 5$, $N \geq 2$, the sum \eqref{Suma} takes the form:
\begin{equation}
  \psi+\alpha_1 \psi^1 + \dots  \alpha_N \psi^N = \frac{m}{r^{2N}}
\end{equation}
We choose the coupling constant such that $\alpha_{n+1}/\alpha_n = \alpha$. Thus, the last equation reads:
\begin{align}
\alpha^0  \psi+\alpha^1 \psi^1 + \dots  \alpha^N \psi^N =& \frac{m}{r^{2N}} \\
\frac{\psi(1- \alpha^N \psi^N)}{1- \alpha \psi} =& \frac{m}{r^{2N}} \label{EqParaPsi}
\end{align}
As we can observe in the last equation, a geometric series is formed. Thus, the form of $\mu(r)$ is determined by the algebraic solution of equations \eqref{EqParaPsi} and \eqref{Psi}.

It is worth mentioning that although for EGB it is irrelevant to write the equations of motion as a geometric series, since there is only one higher-curvature term, for the cubic and quartic cases it will be useful to determine the constraints that the parameters $\alpha$ and $m$ must satisfy. In the cubic case, it will help us establish the conditions that must be met for an integrable singularity to exist, associated with a non-divergent core and the absence of an internal horizon. In the quartic case, it will allow us to numerically manipulate the solution using only these two parameters, graphically determining the behavior of the solution as a regular black hole without the presence of an internal horizon.

\section{Criteria for representing an Integrable singularity in Lovelock Gravity in $d = 2N + 1$ Dimensions} \label{Criterios}

Before defining the structure of the solutions for the cases of EGB and cubic gravity, this section outlines some generic characteristics for representing an integrable central singularity in Lovelock gravity in $d = 2N + 1$ dimensions. Due to the fact that we will show below that in our case this type of solutions is spherically symmetric, in this section we will restrict the analysis to $\gamma = 1$. These characteristics are as follows:

\begin{enumerate}
    \item Finite Non dS core :

    We can note that in the critical case $d-1 = 2N$, that is, in a finite series, the solution behaves as 
\begin{equation} \label{MuNucleo}
  \mu \approx 1 - \left( C_1 \right)^{1/N}, 
\end{equation}
meaning it is finite at the radial origin. Thus, we can observe that near the origin, the core does not correspond to either an (A)dS or Minkowski space, due to the presence of the dominant term of order $N = (d-1)/2$, with $d$ being odd and $d \geq 5$.

\item Related to the absence of an inner horizon near the origin:

As we know, a dS core is such that near the origin $\mu$ behaves like $\mu \approx 1 - C_2^2 r^2$, being $C_2$ a constant, with a spherically symmetric cross-section. In this case, there is an inner horizon at $r_{in} \approx C_2$ near the origin such that $\mu'(r_{in})<0$ and next $\mu'(r_h)>0$ . However, since in our case the dominant term for small $r$ is a finite number, such that the core is given by equation \eqref{MuNucleo}, there is no inner horizon near the radial origin.

It is worth noting that our analysis differs from that in reference \cite{DiFilippo:2024mwm}, which utilizes an infinite tower of higher-curvature terms in QTG, resulting in a solution with vanishing surface gravity at the inner horizon. This solution features multiple degenerate inner horizons. The reference argues that, in four dimensions, such a condition is necessary to prevent classical instabilities associated with mass inflation at the inner horizon.

\item Divergence of the curvature invariants:

By observing the following expressions corresponding to the Ricci and Kretschmann curvature invariants: 
\begin{equation} \label{RicciInvariante}
    R=-\mu'' - 2(d-2) \frac{\mu'}{r} - (d-2)(d-3) \frac{(\mu - \gamma)}{r^2}
\end{equation}
and
\begin{equation} \label{KreInvariante}
  K=  \left(\mu''\right)^2 + \frac{2(d - 2)}{r^2}\left(\mu'\right)^2 + \frac{2(d - 2)(d - 3)}{r^4}\left(\gamma - \mu\right)^2
\end{equation}
repectively, where $\gamma=1$, we can easily verify that both quantities diverge at $r=0$. Thus, although $\mu$ does not diverge at the origin, the curvature invariants do.

\item Integrability of Ricci scalar: \label{RicciIntegrable}

The higher-dimensional Ricci invariant near the origin behaves as:
\begin{equation}
    R \sim \frac{(d - 2)(d - 3)C}{r^2}
\end{equation}

being $C$ a constant. Thus, we can note that the volume integral $\sim R r^{d-2}$ is finite near the origin for $d \geq 4$, which, as mentioned in the introduction, corresponds to an integrable singularity (the definition implies that the integral of $R$ must be integrable, not that of $K$). This condition is satisfied for our case $d = 2N + 1$ for all values of $N \geq 2$.

\item Finite radial Geodesics:

In our case, the line element corresponding to a $(d-2)$-sphere for $\gamma=1$, is given by:

\begin{align}
     \displaystyle   d\Sigma_{(1)} &= d\theta^2_1 + \sum_{j=2}^{D-2} d\theta^2_j \left ( \prod_{k=1}^{j-1} \sin^2\theta_k  \right) 
\end{align}
with $j = 1, \dots, D - 3$. We consider a radial motion described by $\theta_1 = \theta_2 = \dots = \theta_{D-3} = \pi/2$ constant. Thus, any timelike radial geodesic is given by:
\begin{equation} \label{Geodesica}
    \dot{r}^2 + V_{\text{eff}} = \frac{E^2}{2}
\end{equation}
where
\begin{equation} \label{Potencial}
    V_{\text{eff}} = \frac{1}{2} \mu(r)
\end{equation}
with the dot representing the derivative with respect to an affine parameter $\lambda$. In this way, since in our case $\mu$ has a finite value near the origin, equation \eqref{MuNucleo}, the timelike radial geodesic is regular near the origin.

\item Weak nature of the singularity

The Tipler theorem \cite{Tipler:1977zza} establishes whether any object following its world line will inevitably be crushed by a gravitationally strong singularity. Subsequently, reference \cite{Nolan:2000rn} presented an extension of this theorem, determining when a singularity is considered weak or strong. Specifically, if a radial causal geodesic \textcolor{red}{$\bar{\gamma}$}, without angular momentum, terminates in a deformationally weak central singularity, then along $\gamma$, the value of:
\begin{equation} \label{TiplerNolan}
    \lim_{\lambda \to 0} x(\lambda) = \lim_{\lambda \to 0} r(\lambda) \int_{\lambda_1}^{\lambda} \frac{d\bar{\lambda}}{r(\bar{\lambda})^2}
\end{equation}
is finite and nonzero. Thus, it is established that if the singularity is weak, there exists a $c_0 > 0$ such that:
\begin{equation} \label{Corolario}
    r(\lambda) \sim c_0 \lambda \quad \text{as } \lambda \to 0
\end{equation}
Thus, by comparing Equations \eqref{Geodesica} and \eqref{Potencial}, we can observe that for small $r$, the timelike radial geodesic behaves like Equation \eqref{Corolario}. Consequently, Equation \eqref{TiplerNolan} is finite and nonzero. Therefore, the singularity is of weak nature, and thus, an object would not be spaghettified as it approaches it. Thus, the
radial geodesics can be extended up to this point.

As indicated in reference \cite{Nolan:2000rn}, in the case of a geodesic moving in the radial direction, with $\theta_i = \text{cte}$, i.e., without angular momentum, a singularity is of deformationally strong nature if the factor $x(\lambda)$ is zero or infinity in the limit as the singularity is approached ($\lambda \to 0$). Otherwise, if $x(\lambda \to 0) = \text{finite}$, the singularity is weak. That is, in the case of a strong singularity, an observer falling radially into a black hole would be stretched infinitely along the radial direction, resulting in the spaghettification of the observer. As indicated in reference \cite{Lukash:2013ts}, it is important that the flow respecting the space-time symmetry does not suffer from infinite tidal forces. Thus, according to the definition in reference \cite{Nolan:2000rn}, since our equation \eqref{TiplerNolan} yields a finite value, the singularity is of weak nature, which would not imply spaghettification.

\end{enumerate}

Thus, we can deduce that every solution for $n_{\text{max}} = N = (d-1)/2$, with $d$ being odd and $d \geq 5$, $N \geq 2$, in Lovelock gravity can represent a solution with an integrable central singularity, under which there is neither a potentially unstable dS core nor an inner horizon at that location. Moreover, an object would not undergo spaghettification when approaching the central singularity, because the timelike radial geodesics are extendible up to the singularity, and furthermore, the singularity is of weak nature.

Remarkably, these solutions, as we will see below, correspond to vacuum solutions without the need to introduce specific forms of matter in the energy-momentum tensor, unlike previous studies related to integrable singularities \cite{Ovalle:2023vvu, Lukash:2013ts,Casadio:2023iqt,Estrada:2023dcj}.

\section{BHs solutions with an integrable singularity }

\subsection{Einstein Gauss Bonnet solution for $d=5$}

Solving the equations \eqref{EqParaPsi} and \eqref{Psi} for $N=2$ and $d=5$ in the case of a spherically symmetric transverse section, i.e., with $\gamma=1$

\begin{equation}
   \mu = \frac{r^2 + 2 \alpha \pm \sqrt{r^4 + 4m \alpha}}{2 \alpha}
\end{equation}
where $\alpha$ is the Gauss-Bonnet coupling constant, such that in the limit $\alpha \to 0$ and in the $-$ branch, the solution of General Relativity is recovered, known as the General Relativity branch. Thus, we must consider the negative branch in the last equation in order to recover the General Relativity branch of the Einstein Gauss Bonnet theory. Thus, the core is given by:

\begin{equation} \label{coreEGB}
    \mu \big |_{r \approx 0} \approx 1 - \sqrt{\frac{m}{\alpha}}
\end{equation}
\begin{itemize}
    \item We can note that the solution is not divergent near the origin. Thus, as previously described, a timelike radial geodesic can extend up to the singularity, and the singularity is weak in nature. 
    \item Although $\mu$ is finite near the origin, the Ricci scalar diverges at this point due to the form of the core. This is because our core does not possess a Minkowski or (A)dS structure. The absence of a de Sitter core is also related to the lack of an internal horizon close to the potentially unstable core. Thus, in relation to what was described in point \ref{RicciIntegrable} of the previous section, we have an integrable singularity.
   \item On the other hand, it must be satisfied that $m > \alpha$ for a signature change to occur when crossing the event horizon.
\end{itemize}

The solutions of the equation $\mu=0$ are given by:
\begin{equation}
    r_* = \pm \sqrt{m-\alpha}
\end{equation}
Thus, there is only one positive solution, which represents the event horizon:
\begin{equation}
    r_h = \sqrt{m-\alpha}
\end{equation}

\begin{figure}[h]   
    \centering 
    \includegraphics[scale=.6]{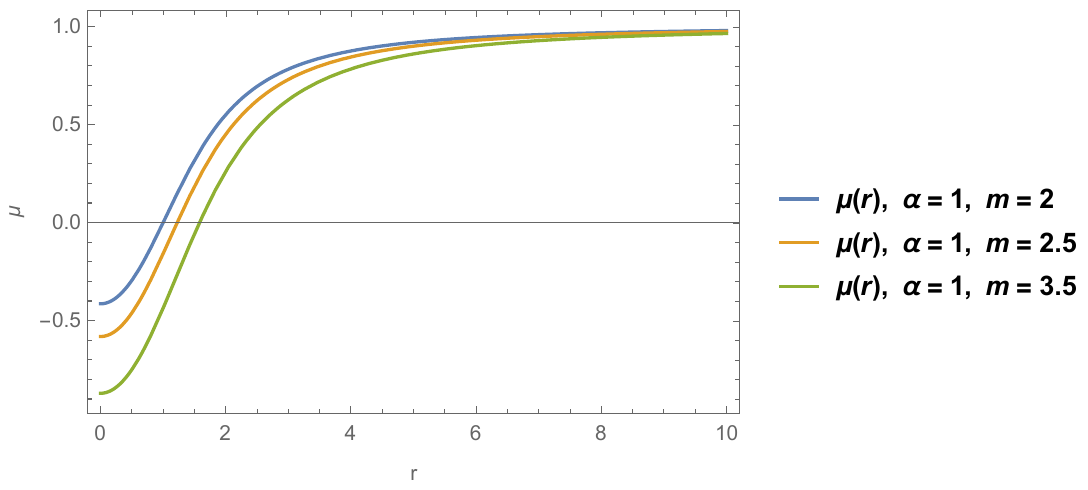} 
    \caption{$\mu$(r) for EGB in $d=5$} \label{FigEGB}
 \end{figure}

Where we can also observe that there is only one event horizon, and no potentially unstable inner horizon exists. This is consistent with the requirement that $m > \alpha$ for the event horizon to have a real value. We can observe the generic behavior of the function $\mu$ in Figure \ref{FigEGB}, where we note that there is no presence of an inner horizon. 

On the other hand, it is worth noting that the solution with $\gamma=-1$ does not represent a black hole geometry, as it lacks an event horizon where the radial coordinate transitions into a time-like coordinate.

\subsection{Cubic solution for $d=7$}

Solving the equations \eqref{EqParaPsi} and \eqref{Psi} for $N=3$ and $d=7$ in the case of a spherically symmetric transverse section, i.e., with $\gamma=1$

\begin{equation}
   \mu =1 + \frac{r^2}{3 \alpha} + \frac{4 \cdot 2^{1/3} r^4 \alpha^2 - \left( 14 r^6 \alpha^3 + 54 m \alpha^4 + 6 \sqrt{3} \sqrt{\alpha^6 \left( 3 r^{12} + 14 m r^6 \alpha + 27 m^2 \alpha^2 \right)} \right)^{2/3}}{6 \alpha^2 \left( 7 r^6 \alpha^3 + 27 m \alpha^4 + 3 \sqrt{3} \sqrt{\alpha^6 \left( 3 r^{12} + 14 m r^6 \alpha + 27 m^2 \alpha^2 \right)} \right)^{1/3}}
\end{equation}

We can also check that the solution of the $\mu=0$ equation is given by the following equation in the parameters space:

\begin{equation}
    0= r_h^4+ \alpha r_h^2 + (\alpha^2-M)
\end{equation}
It is straightforward to check that $m(r_h)$ has a minimum at $m(r_h = 0) = \alpha^2$ and increases monotonically with $r_h$. Therefore, for each value of $r_h$, there is only one corresponding value of $m$, implying the existence of a single event horizon under the condition $m > \alpha^2$. This condition can also be verified by ensuring that the discriminant of the last equation yields a real value.

In this case, the core is given by:

\begin{equation}
    \mu \big |_{r \approx 0} \approx 1 - \left ( \frac{m}{\alpha^2}  \right)^{(1/3)}
\end{equation}

We can again note that the solution remains non-divergent near the origin. The structure of the core is once more different from that of Minkowski or (A)dS. Additionally, the absence of a de Sitter core is related to the lack of an internal horizon close to the potentially unstable core.

We can note that the conditions described in the previous subsection are once again satisfied, namely that timelike radial geodesics can extend up to the singularity, the singularity is weak in nature, the Ricci scalar diverges, and there is the presence of an integrable singularity.

Furthermore, it must be satisfied that $m > \alpha^2$ for a signature change to occur when crossing the event horizon.

The solutions of the equation $\mu=0$ are given by:
\begin{equation}
  r_*= \pm \frac{\sqrt{-\alpha \pm \sqrt{4m - 3\alpha^2}}}{\sqrt{2}}
\end{equation}
equation that, under the aforementioned condition, $m > \alpha^2$, has a unique real and positive solution, which represents the event horizon:
\begin{equation}
    r_h=\frac{\sqrt{-\alpha + \sqrt{4m - 3\alpha^2}}}{\sqrt{2}}
\end{equation}
Where we can also observe that there is only one event horizon, and no potentially unstable inner horizon exists.

\begin{figure}[h]   
    \centering 
    \includegraphics[scale=.6]{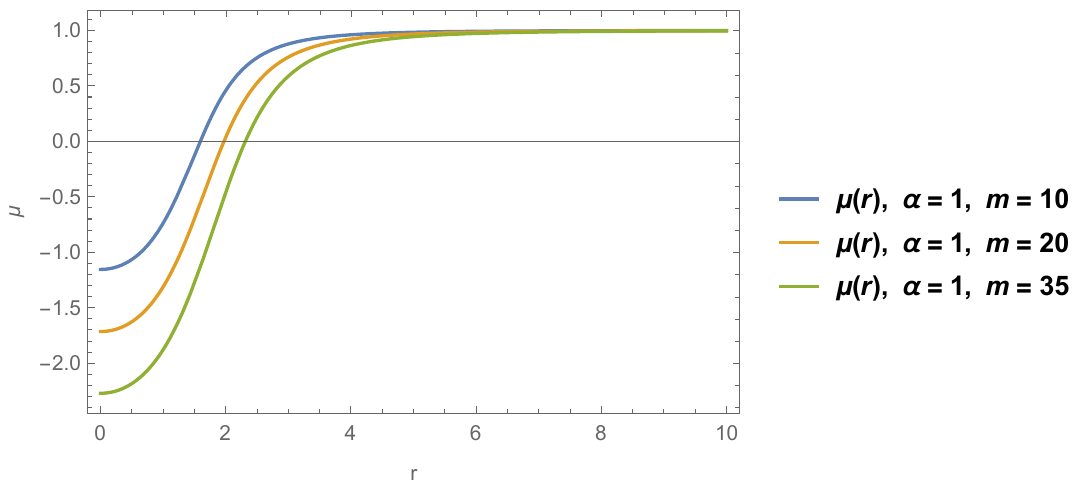} 
    \caption{$\mu$(r) for Cubic case in $d=7$} \label{FigCubic}
 \end{figure}

We can observe the generic behavior of the function $\mu$ in Figure \ref{FigCubic}, where we also note that there is no presence of an inner horizon. 

 On the other hand, it is worth noting that the solution with $\gamma=-1$ does not represent a black hole geometry, as it lacks an event horizon.

 \section{Regular black hole in Quartic gravity for $d=9$}
We solve the equations \eqref{EqParaPsi} and \eqref{Psi} for $N=4$ and $d=9$ in the case of a hyperboloid transverse section, specifically with $\gamma = -1$. In this way, we can verify that the solution capable of representing a black hole with an event horizon is:

\begin{align}
&\mu(r)=\frac{1}{4 \alpha } \Bigg (r^2-4 \alpha + \nonumber \\
      & \frac{\alpha}{\sqrt{3a^3}} \bigg ( - 5 \alpha r^4 - \frac{8 \alpha ^2 \left(6 \alpha  m+r^8\right)}{\sqrt[3]{\frac{3}{2} \sqrt{3} \sqrt{\alpha ^6 \left(256 \alpha ^3 m^3+203 \alpha ^2 m^2 r^8+88 \alpha  m r^{16}+16 r^{24}\right)}+\frac{45}{2} \alpha ^4 m r^4+10 \alpha ^3 r^{12}}} \nonumber \\
&+ 2\ 2^{2/3} \sqrt[3]{3 \sqrt{3} \sqrt{\alpha ^6 \left(256 \alpha ^3 m^3+203 \alpha ^2 m^2 r^8+88 \alpha  m r^{16}+16 r^{24}\right)}+45 \alpha ^4 m r^4+20 \alpha ^3 r^{12}} \bigg)^{1/2} + \nonumber \\
&\sqrt{\frac{2}{3}} \alpha \bigg ( - \frac{5r^4}{\alpha^2} + \frac{4 \left(6 \alpha  m+r^8\right)}{\alpha  \sqrt[3]{\frac{3}{2} \sqrt{3} \sqrt{\alpha ^6 \left(256 \alpha ^3 m^3+203 \alpha ^2 m^2 r^8+88 \alpha  m r^{16}+16 r^{24}\right)}+\frac{45}{2} \alpha ^4 m r^4+10 \alpha ^3 r^{12}}} - \nonumber \\
& - \frac{2^{1/3}}{\alpha^3} \sqrt[3]{3 \sqrt{3} \sqrt{\alpha ^6 \left(256 \alpha ^3 m^3+203 \alpha ^2 m^2 r^8+88 \alpha  m r^{16}+16 r^{24}\right)}+45 \alpha ^4 m r^4+20 \alpha ^3 r^{12}} + \nonumber \\
& \frac{15 \sqrt{3}r^6}{\alpha^{3/2}} \Big ( - 5 \alpha r^4 - \frac{8 \alpha ^2 \left(6 \alpha  m+r^8\right)}{\sqrt[3]{\frac{3}{2} \sqrt{3} \sqrt{\alpha ^6 \left(256 \alpha ^3 m^3+203 \alpha ^2 m^2 r^8+88 \alpha  m r^{16}+16 r^{24}\right)}+\frac{45}{2} \alpha ^4 m r^4+10 \alpha ^3 r^{12}}} \nonumber \\
& + 2\ 2^{2/3} \sqrt[3]{3 \sqrt{3} \sqrt{\alpha ^6 \left(256 \alpha ^3 m^3+203 \alpha ^2 m^2 r^8+88 \alpha  m r^{16}+16 r^{24}\right)}+45 \alpha ^4 m r^4+20 \alpha ^3 r^{12}} \Big)^{-1/2} \bigg )^{1/2} \Bigg )
\end{align}

Expanding for small values of $r$, we find that $C_1 = 0$ in equation \eqref{SolucionBueno}. Therefore, the core is:
\begin{equation}
    \mu \big |_{r \approx 0} \approx - 1 + \frac{r^2}{4 \alpha}
\end{equation}
Since the term of order $r^0$ vanishes in the expansion, the dominant term for this case is the one of order $r^2$.

By substituting the core into the curvature invariants of order \eqref{RicciInvariante} and \eqref{KreInvariante} for \(\gamma = -1\), we can observe that both invariants approach finite values near the origin. Therefore, this solution, under our assumptions, can be interpreted as a regular black hole.

\begin{figure}[h]   
    \centering 
    \includegraphics[scale=.6]{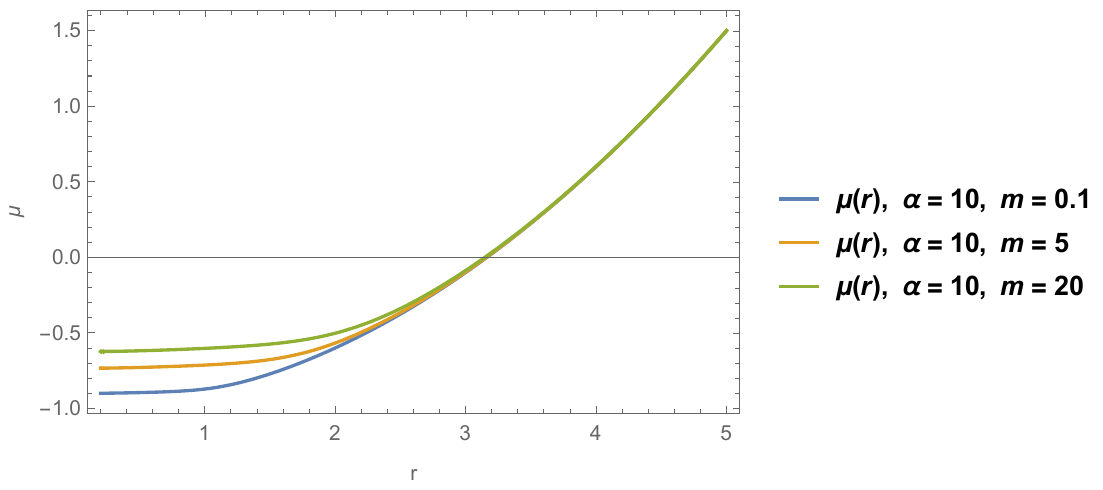} 
    \caption{$\mu$(r) for quartic case in $d=9$} \label{FigQuartic}
 \end{figure}

In the last equation, we can note that the derivative depends on the sign of $\alpha$. Thus, assuming arbitrarily $\alpha > 0$, the function is increasing near the core, indicating that there is no presence of an internal horizon (where the radial coordinate behaves temporally) close to the core . Although it is difficult to analytically describe the global behavior of the solution due to the large number of terms involved, it can be confirmed that $\mu$ is an increasing function. In Figure \ref{FigQuartic}, we observe the general behavior of $\mu$, which indicates that while there is no internal horizon, there is an event horizon.

 \section{Discussion and Summarize}

Typically, both the construction of regular black hole solutions and black holes with an integrable singularity require the inclusion of specific forms of matter in the energy-momentum tensor. Motivated by the recent incorporation of an infinite tower of high-curvature terms in reference \cite{Bueno:2024dgm} to construct vacuum regular black hole (RBH) solutions for quasi-topological gravity, in this work we truncate this tower for odd dimensions $d = 2N + 1$ in Lovelock gravity, also in a vacuum framework. This truncation is also motivated by the fact that, for Lovelock gravity, it is not possible to construct such an infinite tower as $N \to \infty$.

To carry out this work, we first characterized the coupling constants in such a way that the terms present in the equations of motion correspond to a geometric sum, which is useful for determining the conditions that these constants must satisfy in the cubic and quartic cases. In section 3, we have generically determined the criteria that are satisfied for the case $d = 2N + 1$, when the core of the solution is finite but different from Minkowski and AdS: There is an absence of an inner horizon, the curvature invariants diverge, the volume integral of the Ricci scalar is finite (what is known as an integrable singularity), the radial geodesics are finite when reaching the radial origin, and the singularity is weak in nature. These criteria characterize an integrable singularity. Thus, in this work, we have tested the structure of the cores for the cases $d = 5, 7, 9$ to determine whether the solution corresponds to an integrable singularity (or another classification such as regular black holes).

It is worth noting that in the Einstein-Gauss-Bonnet (EGB) case in $d = 5$ and the cubic case in $d = 7$, where there is an integrable singularity, the core of the lapse function behaves non-divergently near the origin. This core is finite but distinct from the typical cores of Minkowski or (A)dS spacetimes. This distinction is related to the absence of an internal horizon near a potentially unstable de Sitter core. The particular form of our core is also associated with the fact that, although the solution is finite at the radial origin, the curvature invariants diverge at this point. As mentioned above, the finiteness of the volume integral of the Ricci scalar is linked to the existence of an integrable singularity rather than a potentially unstable de Sitter core. Remarkably, the timelike radial geodesics can extend up to the singularity, indicating that the singularity is weak in nature. This situation is analogous to that in regular black holes (RBHs), where an object approaching the radial origin does not experience spaghettification. On the other hand, as we have described earlier, the case with $d = 9$ and where the transversal section corresponds to a hyperboloid has an AdS core, so this solution is viewed as an RBH.

Thus, we have determined  new interpretations for vacuum solutions in Lovelock gravity in $d = 2N + 1$. The vacuum Einstein-Gauss-Bonnet (EGB) solution in $d = 5$ and the vacuum cubic solution in $d = 7$ can represent black holes with an integrable singularity, while the vacuum quartic solution in $d = 9$ can represent a regular black hole. For all the mentioned cases, we have determined the conditions that the parameters present in the solutions must satisfy.

As mentioned in the introduction, the theoretical debate regarding whether the presence of an internal horizon close to a de Sitter core is related to instability during inflation has garnered significant attention in recent years. In this regard, it is worth noting that both our solutions representing a black hole with an integrable singularity and the quartic case representing a regular black hole do not exhibit an internal horizon associated with a potentially unstable de Sitter core. Therefore, it would be of interest in the future to investigate the stability of our type of solutions. This would require extensive work, in which the criteria for testing stability would need to be defined, which is beyond the scope of this work.

The fact that it is possible to represent both RBHs and BHs with an integrable singularity without the presence of matter fields in the energy-momentum tensor opens up some questions that could be studied in the future. In this regard, it is well known that the first law of thermodynamics is usually modified by the presence of matter fields in the energy-momentum tensor. In this connection, reference \cite{Estrada:2023dcj} showed a correction proportional to the horizon radius for the entropy. Thus, our vacuum representation of RBHs and BHs with an integrable singularity raises the question of how high-curvature terms would influence the thermodynamics of these vacuum solutions and consequently their evaporation process.

\acknowledgments

Milko Estrada is funded by the FONDECYT Iniciaci\'on Grant 11230247.

\bibliography{mybib.bib}

\end{document}